\documentclass[%
 reprint,
amsmath,amssymb,
 aps,
pre
]{revtex4-2}

\usepackage{graphicx}% Include figure files
\usepackage{dcolumn}% Align table columns on decimal point
\usepackage{bm}% bold math
\usepackage{braket}
\usepackage{mathtools}
\usepackage{tensor}
 \usepackage[verbose]{placeins}

\allowdisplaybreaks

\usepackage[resetlabels,labeled]{multibib}
\newcites{App}{Appendix References}

\begin{document}

\preprint{APS/123-QED}

\newcommand{\sean}[1]{{\color[HTML]{1E8449}#1}}
	
\newcommand{\logit}{{\mathrm{logit}\,} }

\title{Machine learning that predicts well may not learn the correct physical descriptions of glassy systems}% Force line breaks with \\
%%\thanks{A footnote to the article title}%

\author{Arabind Swain}
%\email{arabind.swain@emory.edu}
 \affiliation{Department of Physics, Emory University, Atlanta, GA 30322, USA}%Lines break automatically or can be forced with \\
\author{Sean Alexander Ridout}%
% \email{sridout@emory.edu}
 \affiliation{Department of Physics, Emory University, Atlanta, GA 30322, USA}
 \affiliation{Initiative in Theory and Modeling of Living Systems, Atlanta, GA 30322, USA}
\author{Ilya Nemenman}
%\email{ilya.nemenman@emory.edu}
\affiliation{Department of Physics, Emory University, Atlanta, GA 30322, USA}
\affiliation{Department of Biology, Emory University, Atlanta, GA 30322, USA}
\affiliation{Initiative in Theory and Modeling of Living Systems, Atlanta, GA 30322, USA}

\begin{abstract}
The complexity of glasses makes it challenging to explain their dynamics. Machine Learning (ML) has emerged as a promising pathway for understanding glassy dynamics by linking  their structural features to rearrangement dynamics. Support Vector Machine (SVM) was one of the first methods used to detect such correlations. Specifically, a certain output of SVMs trained to predict dynamics from structure, the distance from the separating hyperplane,  was interpreted as being linearly related to the activation energy for the rearrangement. By numerical analysis of toy models, we explore under which conditions it is possible to infer the energy barrier to rearrangements from the distance to the separating hyperplane. We observe that such successful inference is possible only under very restricted conditions. Typical tests, such as the apparent Arrhenius dependence of the probability of rearrangement on the inferred energy and the temperature, or high cross-validation accuracy do not guarantee success. We propose practical approaches for measuring the quality of the energy inference and for modifying the inferred model to improve the inference, which should be usable in the context of realistic datasets.
\end{abstract}

\maketitle

%\tableofcontents

\section{\label{sec:level1}Introduction}

In recent years, there have been a number of attempts to use Machine Learning (ML) techniques to better understand physical phenomena~\cite{RevModPhysCarleo}. One of the areas that has shown considerable promise is the use of classification algorithms to differentiate between different states of a physical system~\cite{Carrasquilla2017, PhysRevE.95.062122, vanNieuwenburg2017, PhysRevB.94.195105, PhysRevResearch.2.043202, ZHAO2019167938, Schoenholz2016, doi:10.1073/pnas.1610204114, doi:10.1126/science.aai8830, Biroli2020, pollet2019, PhysRevLett.114.108001, Schoenholz2016, 2017pnas, arxiv.2008.09681, Bapst2020, PhysRevLett.127.088007, GIANNETTI2019114639, 2017pnas,2017pnas1, tomi2023}. In some of these cases, ML techniques  manage to go beyond classification, extracting physically interpretable low-dimensional descriptions, such as  order parameters~\cite{Carrasquilla2017, pollet2019,GIANNETTI2019114639}, topological invariants~\cite{PhysRevB.98.085402},  or the energy barriers that determine the rate of rearrangements in a glassy liquid~\cite{Schoenholz2016, 2017pnas, 2017pnas1}. In other words, sometimes ML methods build accurate  {\em physical} models of the studied system, even when the relevant variables describing the physics are  not explicitly in the dataset.  Traditionally, finding such low-dimensional, relevant descriptions requires specialized knowledge, \textit{e.~g.,} of conservation laws. Such successes without this specialized knowledge show the potential of ML techniques to discover new physics with minimal guidance by scientists. However, very little is known about when an ML method, trained to predict a certain aspect of the behavior of a physical system, constructs an accurate physical model, rather than a purely statistical one.

We will answer this question in a simplified, tractable model of the important physical problem of predicting rearrangements of glassy liquids using structural data~\cite{Schoenholz2016,2017pnas,2017pnas1}. Glassy liquids have heterogeneous rearrangement dynamics: in some regions particles rearrange quickly, while others are slow. The degree of heterogeneity, \textit{i.~e.}, the range of dynamical correlations, grows as the temperature is lowered~\cite{Bookberthier}. Despite this, the structural order in a  glass is hard to detect, making the origin of these correlations difficult to understand~\cite{frontiersglassreview, Biroliperspective}.  In recent years, there has been considerable progress in linking the dynamics of glassy liquids to their structure using ML. Support Vector Machines (SVMs)~\cite{PhysRevLett.114.108001, Schoenholz2016, 2017pnas, arxiv.2008.09681, 2017pnas1,tomi2023, seanSVM}, Neural Networks~\cite{Bapst2020, Jung1, Jung2,pezzicoli2023rotationequivariant, GNN1, Alkemade2}, and linear regression~\cite{PhysRevLett.127.088007,AlkemadeJCP, Alkemade2} have been trained on large data sets generated through simulations. Local structural features were used to predict whether a particle rearranges in a specific time period $\Delta t$. All of these methods were shown to predict rearrangements with high accuracy.  The classifiers could also predict rearrangements when applied to data from previously unseen temperatures. Thus, the classifiers learn  local structural predictors of dynamics that generalize across temperatures. In the linear SVM case, the distance to the separating hyperplane, named softness $S$~\cite{Schoenholz2016}, has a simple interpretation as a local energy barrier to rearrangement $\Delta E{\left(S\right)}$.  This is because the probability for a particle to rearrange in some unit time $\Delta t$ given $S$ was numerically found to obey the Arrhenius law,
\begin{equation}
  P{\left(R|S\right)}\propto\exp\left[\Sigma{\left(S\right)}-\Delta E{\left(S\right)}/T\right], 
  \label{eq:arr}
\end{equation}  
which is precisely the probability of rearrangement for a process that requires crossing a single energy barrier $\Delta E{\left(S\right)}$. In particular, $\Sigma{\left(S\right)}$ and $\Delta E{\left(S\right)}$ were found to be linear in $S$. Therefore, this simple linear classifier seems to have learned a physical description of the system, without being instructed to infer it. This learned description has even been used to produce simplified dynamical models~\cite{PRRZhang, PNASZhang, Ridout_2023}. However, there has been no explicit study showing if the success in making predictions signifies that the inferred physical description agrees with the true one. Understanding when the two match is the goal of this work. Specifically, assuming that  there exists an underlying structural variable $S$ such that Eq.~(\ref{eq:arr}) holds in a glassy liquid, we will explore when an SVM can learn the correct variable $S$. 
We focus on  SVMs~\cite{vapnik1995support} (and, more specifically, linear SVMs) in our study because SVMs are interpretable, their performance compares well to other  methods for this system, and the interpretation of statistical properties of the classifier (softness) as a physical quantity (linearly proportional to the Arrhenius energy barrier) was made for SVMs, and not other ML methods.

%There have been studies that have compared the performance of different methods in predicting glassy dynamics~\cite{doi:10.1063/5.0088581,doi:10.1063/5.0128265} but do not shed light on why the ML-based methods work. 

We devise a toy model where a true energy barrier, $\Delta E (\Vec{x})$ describes the probability for a given configuration $\Vec{x}$ to rearrange. We show numerically how the choice of structural variables given to the SVM affects the prediction accuracy and the ability of the trained model to predict the true energy barrier. We show that, if the SVM is given  as the input only those features that contribute linearly 
to  $\Delta E (\Vec{x}) $, then the inferred softness (distance to the separating hyperplane), indeed, predicts the true $\Delta E (\Vec{x})$. This is true even when the SVM is only trained to predict rearrangements, rather than $\Delta E (\Vec{x})$ explicitly. However, we also show that, with a finite amount of training data, the energy barrier estimated through the softness inferred by the SVM can be strongly biased. Surprisingly, this is true even if the quality of prediction, measured by common statistical tests, such  as cross-validation, is high.  Thus, SVM does not necessarily learn the correct energy barriers, even when it seems that it does or should. Since, in real systems, structural variables determining the energy barrier are typically unknown, it is a common practice to provide ML algorithms with a large set of putative predictors of the rearrangement probability. One then hopes that the machine distinguishes the features that directly contribute to the barrier height from those that are correlated with them, and from those that are irrelevant for the prediction. In this scenario, we show that the SVM becomes confused, so that its softness cannot be interpreted as the barrier in the presence of additional features correlated with components of the true energy function.  As a result, applications of SVMs to the inference of the energy of glassy systems will likely fail for many real physical system. Finally, we demonstrate methods to diagnose these problems and to fix them by systematic pruning of the structural features used to predict rearrangements.

\section{\label{sec:level2} Model and Simulations}

We study a toy model, which still contains many of the features relevant for our analysis. In the previous work, Ref.~\cite{Schoenholz2016}, an SVM was used to identify a linear combination  $S_i =S{\left(\vec{x}_i\right)}= \sum_{j=1}^n  \alpha^j x^j_i$ of structural features $\Vec{x}_i$, associated with a specific particle $i$, such that the probability of rearrangement for the particle is as in Eq.~(\ref{eq:arr}). Specifically, in order to reproduce Eq.~(\ref{eq:arr}), we require a model where (i) each particle $i$ is described by $n$ structural variables $\vec{x}_i =\{x^1_i,x^2_i,\dots,x^n_i\}$, which vary among the particles; (ii) each particle has a rearrangement energy barrier $\Delta E{\left(\Vec{x}_i\right)}$, and (iii) the probability to rearrange depends on $T$ and $\Delta E{\left(\Vec{x}_i\right)}$ with a law that tends to the Arrhenius law for low temperatures. The simplest model with these properties is one where all $n$ dimensions of $\Vec{x}_i$ are drawn independently at random, and the true energy barrier is a linear function of the $n$-dimensional $\Vec{x}_i$. Thus, for each particle  $i=1,\dots, N$, we generate an $n$-dimensional coordinate vector $\vec{x}_i =\{x^1_i,x^2_i,\dots,x^n_i\}$ as  
\begin{equation}
    x^j_i \sim \mathcal N(0,\sigma^2) \quad \forall \quad j=1,\ldots,n \quad \textrm{and} \quad i=1,\dots,N.
\end{equation}
We then assume that the energy barrier to rearrangement is a linear combination of these coordinates
\begin{equation}
    \Delta E(\vec{x}_i)=\sum^{n}_{j=1} \alpha^j x_i^j.
    \label{eq:energy}
\end{equation}
Finally, for each configuration, we determine whether or not it rearranges by sampling a binary random variable $R_i=\pm 1$ (where $\pm1$ stands for presence/absence of a  rearrangement) from
\begin{equation}\label{eqsigmoid}
    P(R_i=1 \mid \vec{x}_i)=\frac{e^{-\beta \Delta E(\vec x_i)}}{1+e^{-\beta \Delta E(\vec{x}_i)}},
\end{equation}
which reduces to the Arrhenius form at low $T$ while remaining below $1$ at high $T$.

We then train a linear SVM \cite{vapnik1997} to predict $R_i$ from $\vec{x}_i$, for all $i=1,\dots,N$. As is the common practice, for the training, we standardize all $x$s to have zero mean and unit variance. Thus,  drawing $x^j$ from $\mathcal{N}{(0,\sigma^2)}$ is equivalent to drawing them from $\mathcal{N}{(0,1)}$ and absorbing the standard deviation into the definition of $\alpha$, which is what we do. Further, the results shown below are all evaluated at $\alpha^j=1.2$, and we verified separately that this choice does not change our conclusions qualitatively (not shown). 

After training the SVM, we define the {\em softness} $S_i$ for state $\vec{x}_i$ as the signed distance to the separating hyperplane, as in previous work~\cite{Schoenholz2016}. We then want to estimate the probability of rearrangement $P(R|S)$, to see if the softness defines it well. In Ref.~\cite{Schoenholz2016}, this probability was estimated as the frequency of rearrangements in a certain small bin of  $S$. Instead, to remove artifacts caused by the finite bin width, we estimate $P(R|S)$ using a logistic regression model.
\begin{figure*}[htb!]
    \centering
    \includegraphics[width=0.8\textwidth]{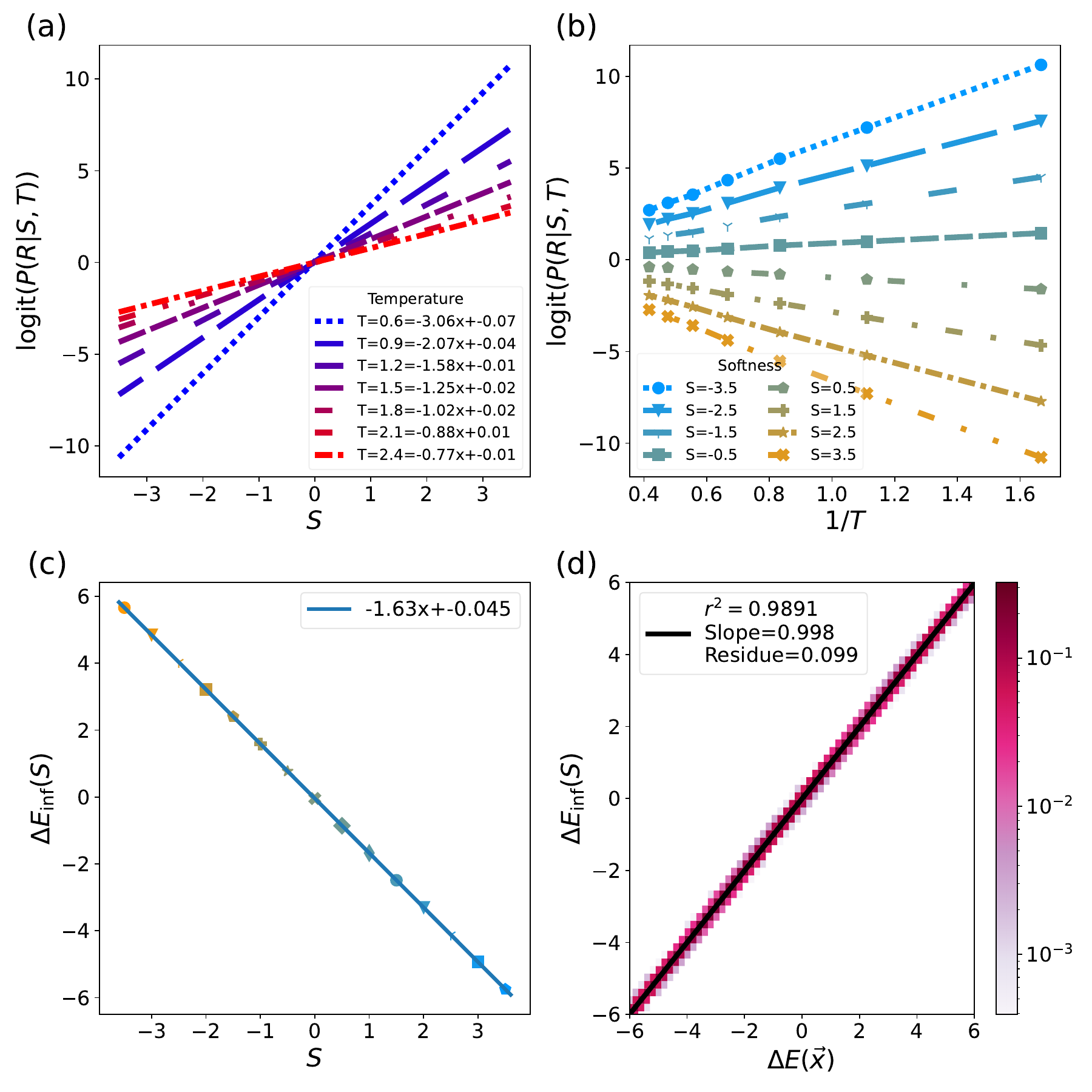}
    \caption{\textbf{Relationship between softness $S$ and $\Delta E (\vec x)$ for symmetric distribution of training energies for a large training set size, $N=10^6$.} \textbf{(a)} $\logit P\left(R|S\right)$ derived from fitting the logistic  curve to the probability of rearrangement as a function of $S$ for different temperatures $T$. \textbf{(b)} $\logit P{\left(R|S,T\right)}$ vs.\ 1/$T$ for 15 different values of softness. \textbf{(c)} The inferred $\Delta E_{\mathrm{inf}}(S)$, calculated from $\logit P{\left(R|S,T\right)}$, as a function of $S$. % It is generated by calculating the slope of $\logit P{\left(R|S,T\right)}$ vs 1/$T$ for each softness.
    \textbf{(d)} 2D joint density plot and the linear fit of the true energy barrier $\Delta E(\Vec{x})$ vs.\ the inferred energy barrier $\Delta E_{\mathrm{inf}}(S)$ (we plot the joint density instead of the scatter for clarity of the visualisation).}
    \label{fig:infinitedatalimit}
\end{figure*}

In a glass, energy barriers should be strictly positive, and the probability for a typical particle to rearrange is tiny. To remove biases in the inference, one typically balances the dataset used for training to have similar numbers of particles that do and do not rearrange~\cite{Schoenholz2016}. In our model, Eq.~(\ref{eq:energy}), we achieve this balance by explicitly centering $\Delta E$ at zero. We checked numerically that this choice does not affect our conclusions qualitatively.

A large number of structural features are used to train an SVM to predict glassy dynamics~\cite{Schoenholz2016}. These features, however, are correlated. To observe the effect of these correlations on the ability of the SVM to predict the correct energy, for some of our simulations, we give as input to the SVM a $2n$-dimensional coordinate vector $\left(\vec{x}_i, \vec{z}_i\right)$, where  $z_i^j  = (x_i^{j})^2 \sum_{j_1}  x_i^{j_1}$, and all $x$'s remain uncorrelated, as before.  There is nothing particular about this choice of additional variables $z^j$ correlated with $x^j$, besides that we wanted to preserve the same symmetry under parity (even order contributions would average out for symmetric $x$s). Further, we wanted these spurious extra dimensions to be non-linearly correlated with $x$s, modeling nonlinear correlations between values of different radial and angular density functions in \cite{Schoenholz2016}. We believe that our conclusions will be qualitatively the same for other choices of spurious correlated variables obeying these conditions.  We then train the SVM to predict rearrangements from this expanded set of coordinates and evaluate the effect of the correlated input variables on the quality of the model the SVM builds.

\begin{figure*}
    \centering
    \includegraphics[width=0.8\textwidth]{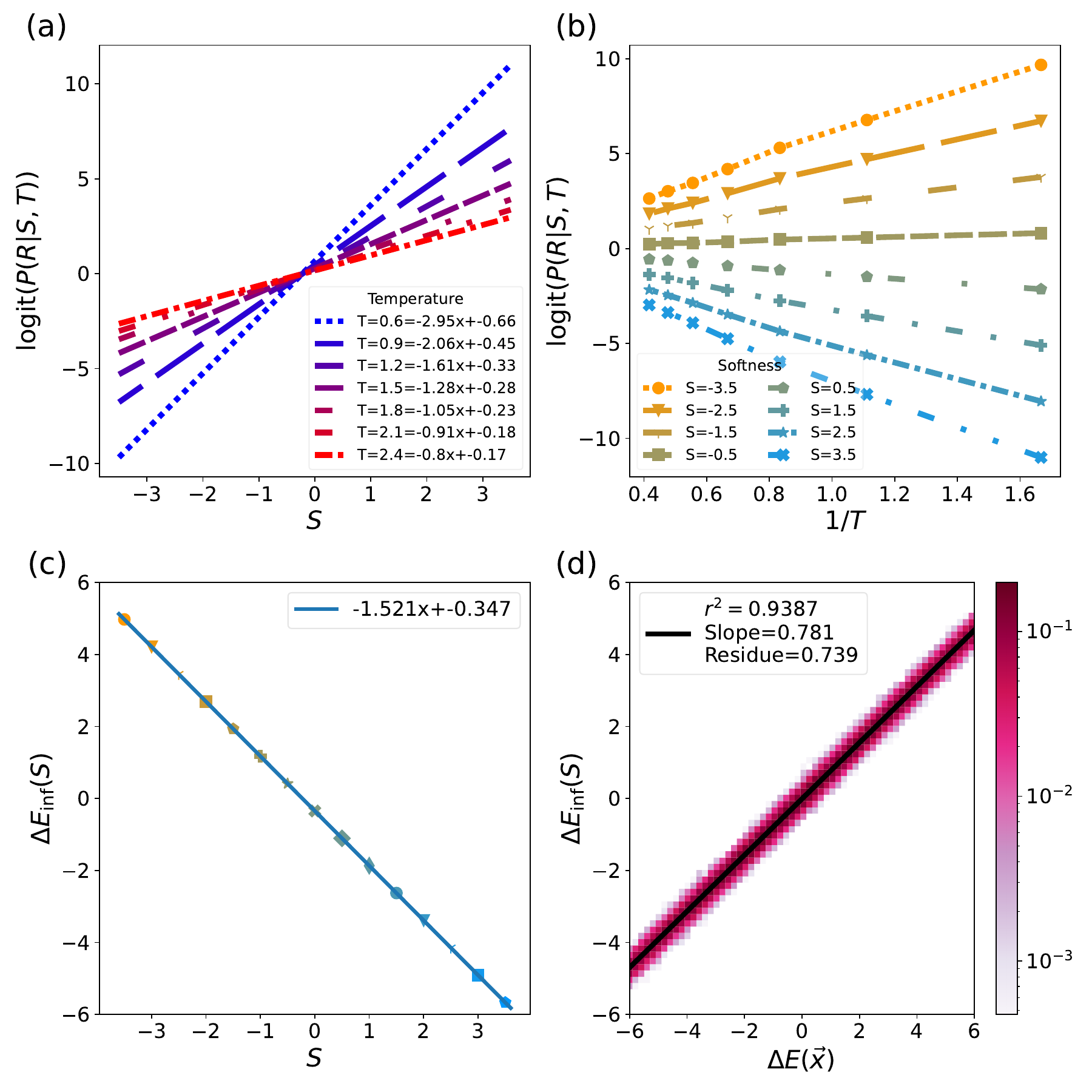}
    \caption{\label{fig:errorslopeprediction}\textbf{Relationship between softness and $\Delta E (\vec x)$ for symmetric distribution of training energies for a small training set size, $N=10^3$.} \textbf{(a), (b), (c)} Same as in Fig.~\ref{fig:infinitedatalimit}. In \textbf{(d)} The true energy barrier $\Delta E(\Vec{x})$ vs the inferred energy barrier from SVM $\Delta E_{\mathrm{inf}}(S)$ is plotted. Note that, to the extent that the slope in \textbf{(d)} is not 1, the correct energy is not learned. }
    
\end{figure*}

\section{\label{sec:level3} Linear SVM  can learn the true energy barrier in the infinite data limit}

First, we test whether or not the softness $S$, inferred by the SVM from a very large sample, is a good approximation for  $\Delta E{\left(\vec{x}\right)}$ from Eq.~(\ref{eq:energy}). We use $N=10^6$ training samples with $5\times 10^5$ examples each of rearranging and non-rearranging configurations to train the SVM. The distribution of energies in the training sample is symmetric. We have 14 independently sampled input dimensions, with $\alpha^j = 1.2$ for $j=1,\dots, 10$ and $\alpha^j = 0$ for $j=11, \dots, 14$. Thus, 10 dimensions determine the energy, while the other 4 dimensions can be seen as Gaussian noise  uncorrelated with  any of the relevant input dimensions.

In Fig.~\ref{fig:infinitedatalimit}(a), we show the relationship between the probability for particles to rearrange, $P{\left(R|S\right)}$, and $S$ by plotting $\logit P{\left(R|S\right)}\equiv \log [P(R|S)/(1-P(R|S)]$) vs.\ $S$. 
$P{\left(R|S\right)}$ is calculated by fitting a logistic regression that predicts whether a particle is rearranging from its $S$. This plot is analogous to the $\log P{\left(R|S\right)}$ vs.\ $S$ plots in earlier studies~\cite{Schoenholz2016} since in our model $\logit P\left(R| \Delta E\right)$ is linear in $\Delta E$. The plot shows a similar linear relationship between $\logit  P{\left(R|S\right)}$ and $S$. When $\logit {P{\left(R|S\right)}}$ is plotted as a function of $1/T$ for several values of softness (Fig.~\ref{fig:infinitedatalimit}b), we also see a linear relationship between $\logit P{\left(R|S\right)}$ and $1/T$  as observed in earlier studies~\cite{Schoenholz2016}. As in the previous work~\cite{Schoenholz2016}, the slope of $\logit P{\left(R|S\right)}$ vs.\ $1/T$ for each softness $S$ is used to infer the corresponding energy barrier  $\Delta E_{\mathrm{inf}}(S)$ in Fig.~\ref{fig:infinitedatalimit}c. This $\Delta E_{\mathrm{inf}}(S)$ is analogous to the barrier energy $\Delta E{\left(S\right)}$ in the Arrhenius rate equation, Eq.~\ref{eq:arr}. As one can see, the inferred barrier energy, $\Delta E_{\mathrm{inf}}(S)$, has a linear relationship with softness, $S$. Thus, our model, in this limit, reproduces the observations of previous work~\cite{Schoenholz2016}: the probability of rearrangement is exponential in the distance $S$ to the separating hyperplane, a.k.a.\ softness, and this distance has an interpretation as an inferred energy barrier  $\Delta E_{\mathrm{inf}}(S)$.

Unlike in past work, in our model, the true energy barriers are \textit{known}. Thus, we then can compare the inferred energy barrier $\Delta E_{\mathrm{inf}}(S)$ to the true energy barrier $\Delta E(\Vec{x})$ for each configuration  $\vec{x}_i$ in the test set. We plot the inferred energy vs. the true energy, as well as a linear regression line between the two in Fig.~\ref{fig:infinitedatalimit}(d). Since the slope of the fit is $\approx 1.0$, and the scatter around the linear fit is small, we conclude that the SVM, indeed, learns the real energy barrier $\Delta E(\Vec{x})$ with a high degree of accuracy. We also find that the SVM captures the real energy when trained on unsymmetrical data where all energy barriers are positive (Appendix \ref{appendunsymmetric}).

\section{\label{sec:level4}Large training sets are required for SVM to learn true energy barriers}

\begin{figure}
    \centering
    \includegraphics[width=0.445\textwidth]{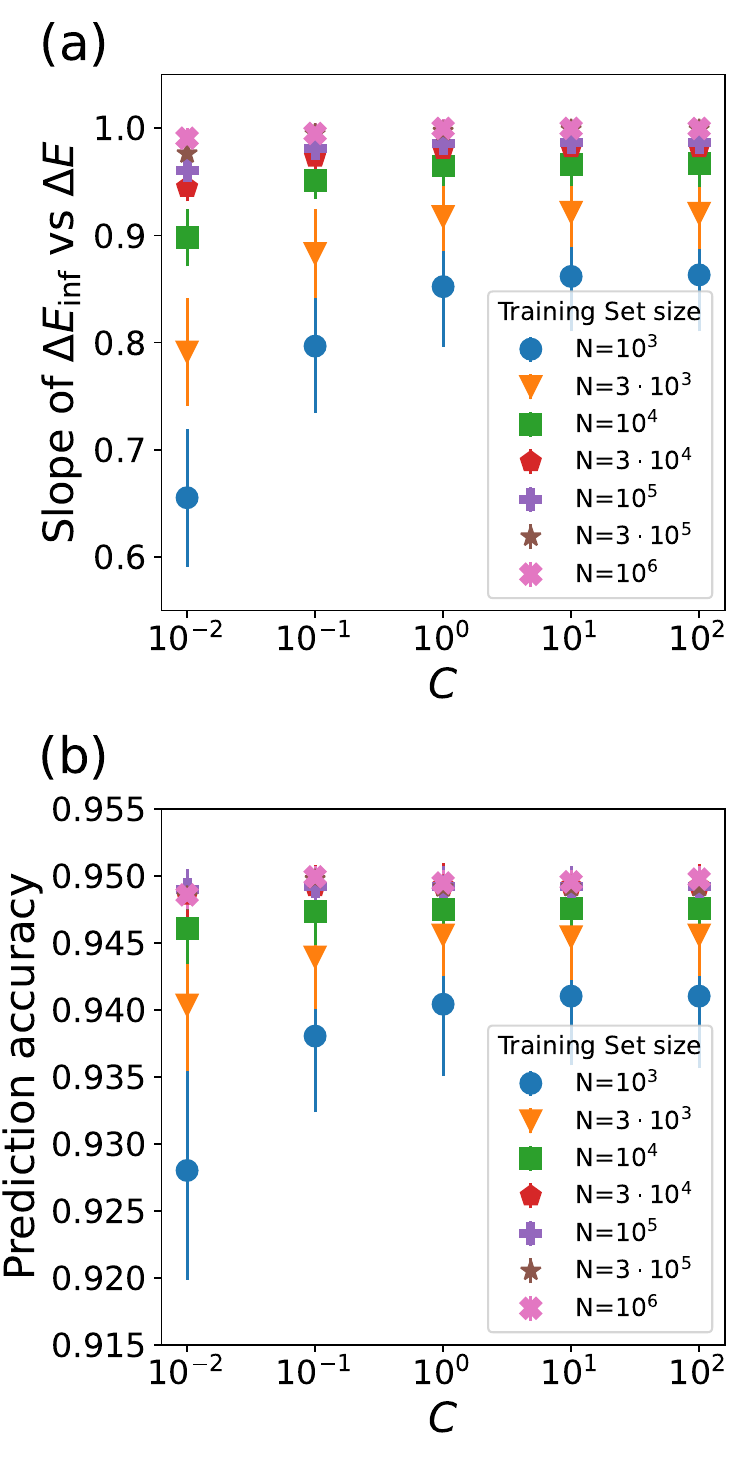}
    \caption{\label{fig:SlopeandPsizedependence}Slope of inferred energy {\bf{(a)}} $\Delta E_{\mathrm{inf}}(S)$ vs.\ real energy $\Delta E(\Vec{x})$ and the prediction accuracy {\bf{(b)}} for different sizes of training data as a function of the SVM cost parameter $C$. The training and test data were generated at $T=0.4$.}
\end{figure}
For real-world problems, we do not have access to an infinite (extremely large) amount of data. Thus, it is natural to ask whether inferred energies are still accurate for smaller training sets. For this,  we repeated the analysis of section~\ref{sec:level3} with varied training set size $N= 10^{3},\dots, 10^6$.

As shown in  Fig.~\ref{fig:errorslopeprediction}(a--c), when $N=10^3$, the inference procedure still seems to work. That is, $\logit P{\left(R|S, T\right)}$ is still a linear function of $S$, and it still appears to be linear in $1/T$. This  allows us again to infer the energy barrier $\Delta E_{\mathrm{inf}}{\left(S\right)}$, which is linear in $S$. However, regressing $\Delta E_{\mathrm{inf}}{\left(S\right)}$ against the true $\Delta E{\left(\vec{x}\right)}$ shows that the inferred energy is \textit{biased}, consistently underestimating the magnitude of the true energy by nearly $15\%$. Since the variance of the true energy is a sum of the variance explained by $S$ and the variance unexplained by $S$, the error must always have this sign: if the energy is inferred incorrectly, its variance will be underestimated. This point is discussed further in Section \ref{sec:level5}.

Fig.~\ref{fig:SlopeandPsizedependence}(a) show how this underestimation depends on $N$. Further, Fig.~\ref{fig:SlopeandPsizedependence}(b) shows the $N$ dependence of the classification (rearranged or not) prediction accuracy of our fitted model on a test set, different from the training one.  To verify that fitting and prediction errors do not come from suboptimal choices during training, in this Figure, we also change the value of the SVM training hyperparameter $C$, which controls when the SVM treats data points that are labeled differently from their neighbors as outliers vs.\ true data that should be fitted. For small $N$, regardless of $C$, the true energy is underestimated. For large $N$, the quality of the fits improves, and the prediction accuracy as well as the error in slope become largely insensitive to $C$.

In practice, the true energy is rarely known. Thus detection of the bias shown in Figs.~\ref{fig:errorslopeprediction}(d), \ref{fig:SlopeandPsizedependence} is nontrivial in experimental applications. Indeed, simple checks, such as verifying the linearity of plots in Fig.~\ref{fig:errorslopeprediction}(a,b,c), do not reveal this error. Further, the underestimation of the barrier magnitude is also difficult to diagnose  by looking at the prediction accuracy, Fig.~\ref{fig:SlopeandPsizedependence}(b). When the true energy is underestimated by $15\%$, the prediction accuracy is still $94\%$ ($C=10^2$, $N=10^3$), which is only $1\%$ lower than the highest value obtained with large $N$. Since we do not have any prior information about the maximum possible prediction accuracy for specific experimental data sets, these figures suggest that, judging by the prediction accuracy only, one can never be sure if the learned energy is a good estimate of the true one: a seemingly high accuracy is not enough!

\begin{figure*}
    \centering
    \includegraphics[width=0.8\textwidth]{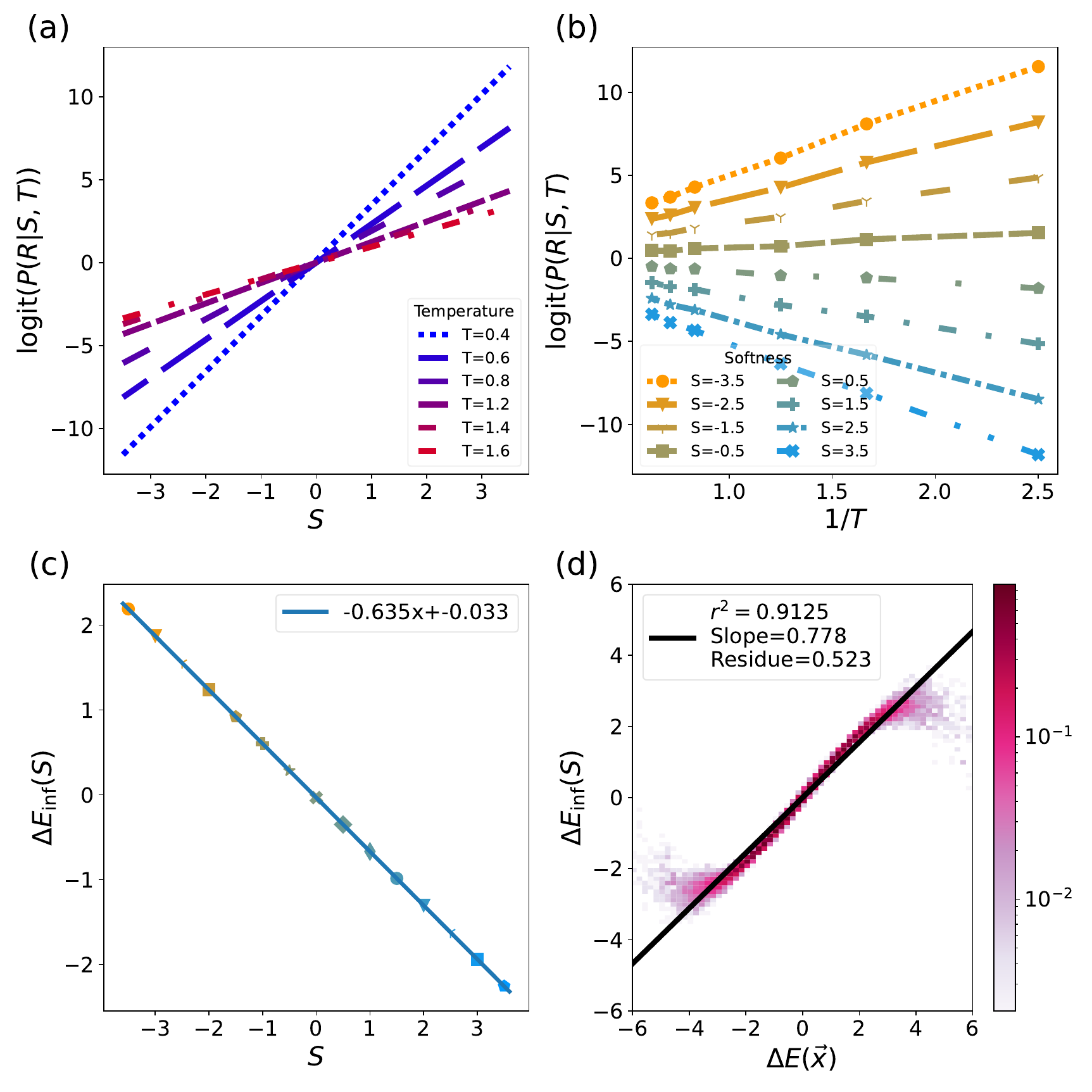}
    \caption{\textbf{Relationship between softness and $\Delta E (\vec x)$ for a symmetric distribution of training energies and with spurious, correlated input terms.} Same plotting convention are used as in Figs.~\ref{fig:infinitedatalimit}. In \textbf{(d)} the true energy barrier $\Delta E(\Vec{x})$ vs the inferred energy barrier from SVM $\Delta E_{\mathrm{inf}}(S)$ is plotted. The error to the fit is given by the purple semi transparent spread on both sides of the fit on a 2D density plot. Note that, to the extent that the slope in \textbf{(d)} is not 1, the correct energy is not learned. Also the deviation between the fit and the 2D density plot at the edges shows that even though a linear fit was used to fit the energy and softness and it fit has a high $r^2$ value the underlying function one is trying to fit is not really linear in $S$.}
    \label{fig:cubicnotworking}
\end{figure*}

\section{\label{sec:level5}Presence of redundant features in the input data degrades the quality of the inference }

\begin{figure}
    \centering
    \includegraphics[width=0.445\textwidth]{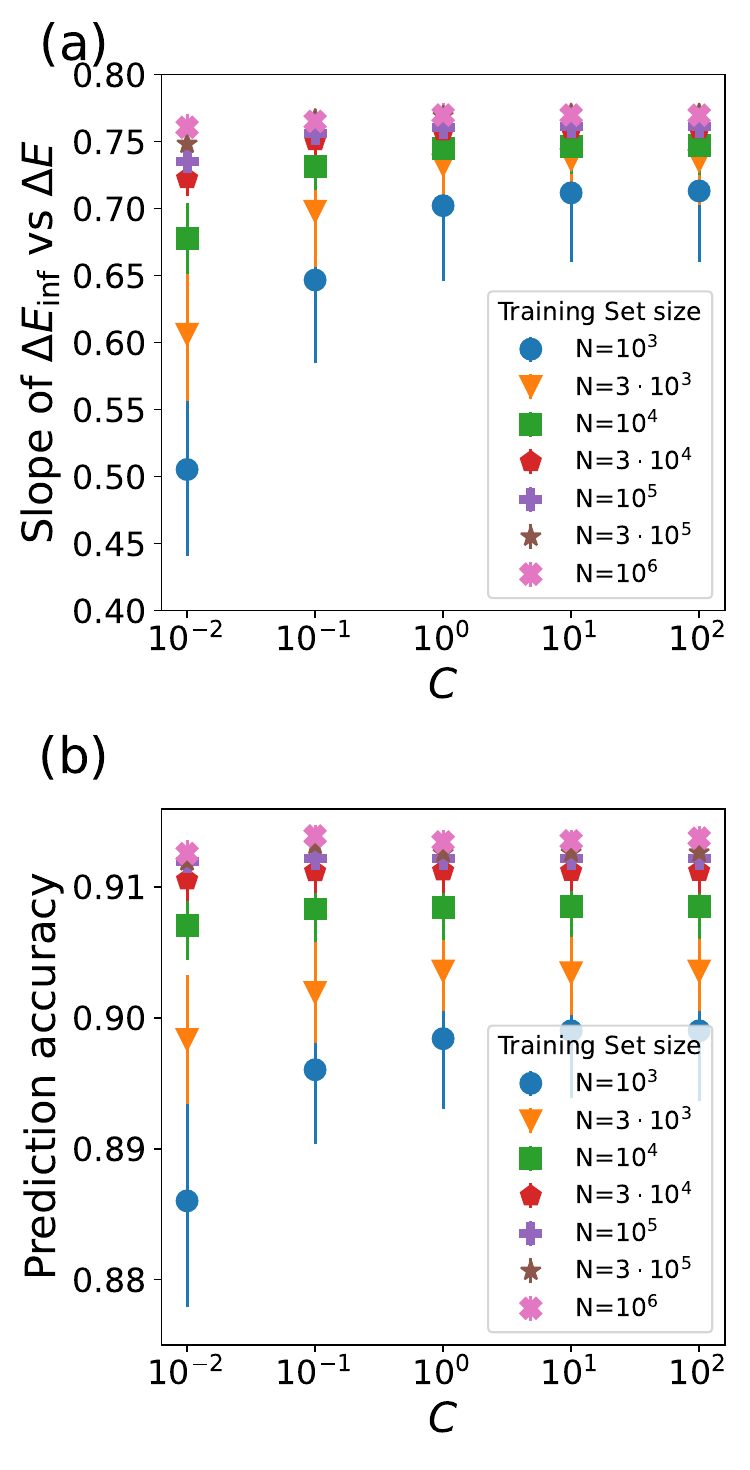}
    \caption{{\bf (a)} Slope of the inferred energy, $\Delta E_{\mathrm{inf}}(S)$, vs.\ the true energy, $\Delta E(\Vec{x})$, and {\bf (b)} the prediction accuracy for the model with spurious correlated inputs. Same plotting conventions as in Fig.~\ref{fig:SlopeandPsizedependence}.}
    \label{fig:sizedependencecubic}
\end{figure}

In Ref.~\cite{Schoenholz2016}, 166 inputs were used for predicting rearrangements. However, many of these inputs were correlated with one another. To model this, we repeat our analysis using a higher-dimensional input vector. For this, as explained in Sec.~\ref{sec:level2}, we train the SVM on a 20 dimensional input. Of these input dimensions, $x^j_i$, $j=1,\dots,10$ were independently sampled from a Gaussian distribution, and the remaining inputs were strongly nonlinearly correlated with them.  We again train an SVM on $N=10^6$ balanced data points. The $\logit P{\left(R|S\right)}$ vs.\ $S$ plot (Fig.~\ref{fig:cubicnotworking}a), $\logit P{\left(R|S,T\right)}$ vs.\ 1/T plot (Fig.~\ref{fig:cubicnotworking}b) and the inferred energy $\Delta E_{\mathrm{inf}}(S)$ vs softness plot (Fig.~\ref{fig:cubicnotworking}c) again are linear, as in Fig.~\ref{fig:infinitedatalimit} and the previous work~\cite{Schoenholz2016}. However, plotting the inferred energy $\Delta E_{\mathrm{inf}}(S)$ vs.\ the true energy barrier $\Delta E(\Vec{x})$ for each configuration and producing a linear fit between them, cf.~Fig.~\ref{fig:cubicnotworking}d, we see that the magnitude of the inferred energy is underestimated compared to the true energy even for very large $N$ (cf.~Fig.~\ref{fig:sizedependencecubic}).  Looking at the optimal hyperplane learned by the SVM, we observe that the hyperplane contains contributions from the input variables that do not contribute to the true energy (not shown). One would not be aware of this problem from Fig.~\ref{fig:cubicnotworking}(a-c) alone. We remind the reader that the true energy needed to produce Fig.~\ref{fig:cubicnotworking}(d)  is typically unknown.   

\begin{figure}
    \centering
    \includegraphics[width=0.445\textwidth]{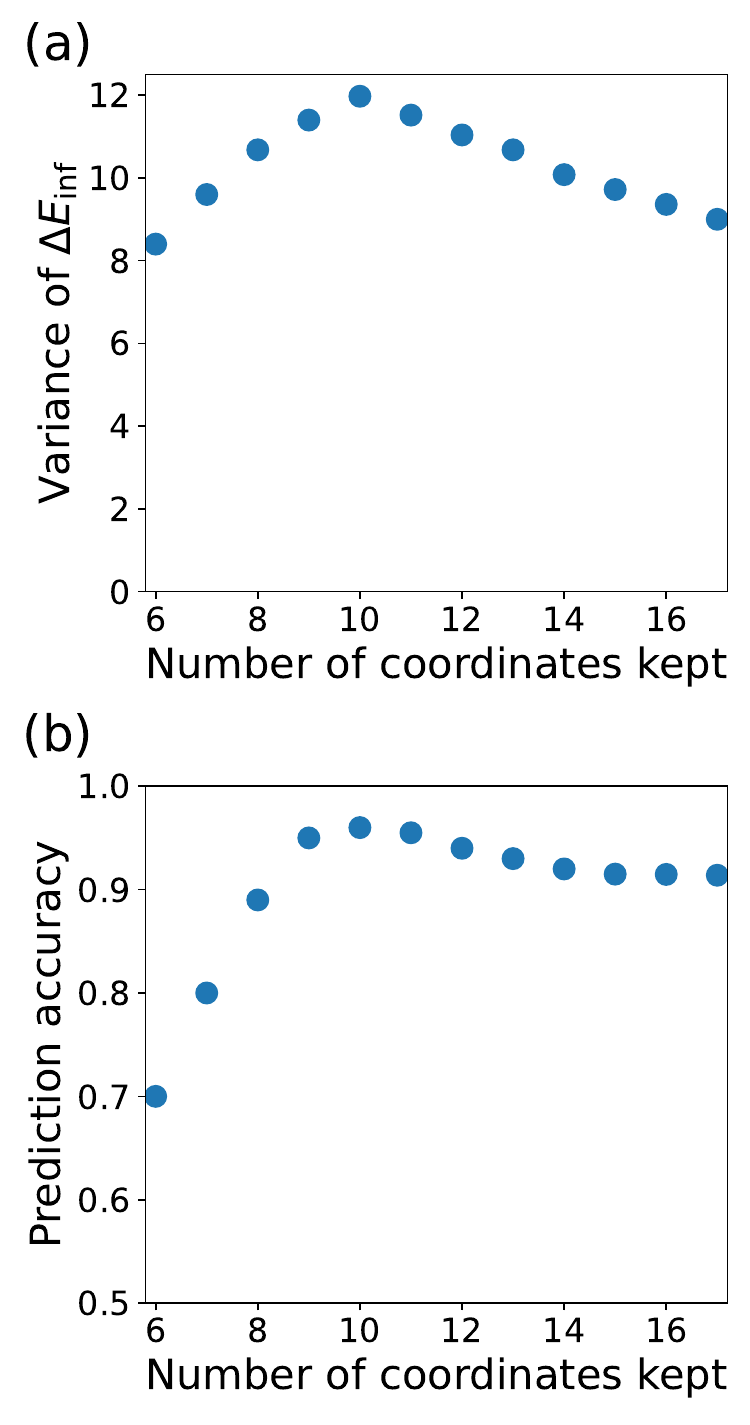}
    \caption{Plot of variance of $\Delta E_{\mathrm{inf}}$ and the prediction accuracy as a function of number of coordinates kept. The variance of $\Delta E_{\mathrm{inf}}$ is more sensitive for detecting relevant dimensions. The inferred energy variance at the peak  matches well the true energy variance ($12$ in our units). }
    \label{fig:parametertuning}
\end{figure}

To design a method for identifying the bias from data, we note that the true variance in the energy is a sum of the variance explained by $S$ and the variance conditional on  $S$.  Thus, if we can find a different set of coordinates that allows the SVM to learn a different $S$ that is \textit{closer} to the true energy, this improvement should manifest as an increase in the variance of $\Delta E_{\mathrm{inf}}$. Our approach is then  to reduce dimensionality of the input space, aiming to remove the correlated dimensions and increase the accuracy of the model at the same time. A particular version of this approach is known in the SVM literature as Recursive Feature Elimination (RFE)~\cite{RFE2002} procedure. RFE has been used in earlier work on predicting rearrangements~\cite{Harrington2019, Ivancic} for pruning the dimensionality of SVM inputs. Assuming that all input dimensions are normalized to the same variance, RFE works by removing the input dimension with the smallest magnitude contribution to the separating hyperplane. One then refits the SVM and continues the process iteratively. Figure~\ref{fig:parametertuning}a shows the variance of the inferred energy as a function of the number of inputs kept by the RFE procedure. The peak in the variance clearly matches the true number of dimensions that contribute to the energy in our model. Figure~\ref{fig:parametertuning}b shows a corresponding (but broader) peak in the prediction accuracy as well. These analyses bode well for using RFE for pruning the input data and resulting in a more accurate inference of the energy barrier in real world problems.

\section{\label{sec:level6} Discussion}
We have shown that, in our toy model, one can always  use a linear SVM to predict rearrangements with a high accuracy, though the amount of data needed for this might be larger than what typical experiments would allow in realistic cases. However, even if the inference seems successful, the inferred energy barrier matches the true energy only in specific cases. Crucially, by observing a high prediction accuracy or high quality linear relationship between softness, log rearrangement probability, and $1/T$, one  cannot conclude that the correct energy has been learned. The problem becomes severe---even in our simple model---when the input data has extra features, potentially nonlinearly correlated with true variables describing the model. This questions the use of ML methods, and specifically SVMs, for inference of energy barriers in glasses.

For our model, we have demonstrated a method to diagnose and fix this problem: recursive feature elimination (RFE) can be used to remove ``confusing'' input features. By tracking the variance of the inferred energy barriers or of $\log{P{\left(R|S\right)}}$, which is maximal when the true barriers are learned, improvements in the inference of the barriers can be detected, even though the true barriers are unknown and the prediction accuracy may change little. RFE is particularly natural in our problem because there is a clear division between important and unnecessary input dimensions. For other systems, RFE may not be the best method for adjusting the set of input features. For example, if the input features are a discretization of the pair correlation function $g(r)$, it may be more natural to coarsen this discretization, or to change the choice of basis functions, than to eliminate specific input features. However, our criterion for comparing different choices of input features is general would still stand: features that produce a larger variance in the inferred energy barriers should be closer to predicting the true barriers. We expect it to be true in general that the choice of features for the inference will affect whether or not the true energy is learned, so that different possible choices should be compared using this criterion.

In our simple model, the probability of particle rearrangement is purely a function of energy. However, when SVMs are used to predict rearrangements in real systems, the probability is a function of energy as well as of an entropic prefactor, both of which are found to depend on $S$ \cite{Schoenholz2016}, see Eq.~(\ref{eq:arr}). In addition, there are other complications not present in our toy model, such as ambiguity in the identification of rearrangements. We expect such complications to only strengthen our conclusion that a good prediction accuracy does not guarantee that the ML model learns the true values of the  energy barriers.

It may seem surprising that the addition of extra coordinates degrades the prediction accuracy and the quality of inference of $\Delta E_{\mathrm{inf}}$. Conventional wisdom is that such overcomplete representation  should improve SVM accuracy by creating a higher-dimensional embedding space, in which the data become linearly separable~\cite{vapnik1997}. It is possible that the failure of this intuition in our case comes from the probabilistic nature of rearrangements: for any $\vec{x}$, there are both rearranging and non-rearranging examples, at least in the $N\to\infty$ limit. Thus, the data are fundamentally not separable, irrespective of the space in which we embed them. 

The process of adding more correlated coordinates explicitly to our input is similar to using some nonlinear kernel on the original data. SVM kernels allow us to create high dimensional embeddings that are nonlinear functions of the input coordinates without having to explicitly evaluate the embedding. Thus, our results may also explain the observation in~\cite{Schoenholz2016} that, for predicting rearrangements in glasses, linear SVMs had the highest prediction accuracy, while other kernel based methods did no better, and neural networks underperformed.

In our work, we have focused specifically on linear SVMs, rather than other ML methods, because this is the only method, which has been used in the past to explicitly deduce the underlying energy barriers from the inferred statistical model. However, note that we have chosen the true energy function to be expressible by a linear SVM. Further, note that more complex ML methods are generally thought to behave similarly to kernel methods \cite{jacot2018neural,roberts2022principles}. Thus we expect that our results are not caused by the simplicity of linear SVMs, and they will generalize to other ML approaches to the problem of learning energy barriers in glassy systems.

Our  results may have implications for many systems beyond supercooled liquids, for which the underlying ``physics'' must be learned from an ML model trained on the data. Indeed, we have shown that, even given a  powerful ML model that can express the true underlying physics, an arbitrarily large amount of training data, and a good prediction accuracy, the model may fail to learn a correct physical description even in a relatively simple scenario. We suspect that, in real world applications, this problem will become even more severe. One must then use independent methods---going beyond prediction accuracy---to evaluate the model quality.

\begin{acknowledgments}
We thank Andrea Liu, Daniel Sussman, Eric Weeks, Daniel Weissman, and Ahmed Roman for important feedback. This work was funded, in part, by a Simons Foundation Investigator grant and the NIH grant 5-R01-NS084844. We also acknowledge the use of the HyPER C3 cluster of Emory University's AI.Humanity Initiative. 
\end{acknowledgments}

\nocite{*}

\bibliography{apssamp}% Produces the bibliography via BibTeX.

\appendix
\section{SVM learns the real energy even when trained on data with positive energy barriers}
\begin{figure*}
    \centering
    \includegraphics[width=0.8\textwidth]{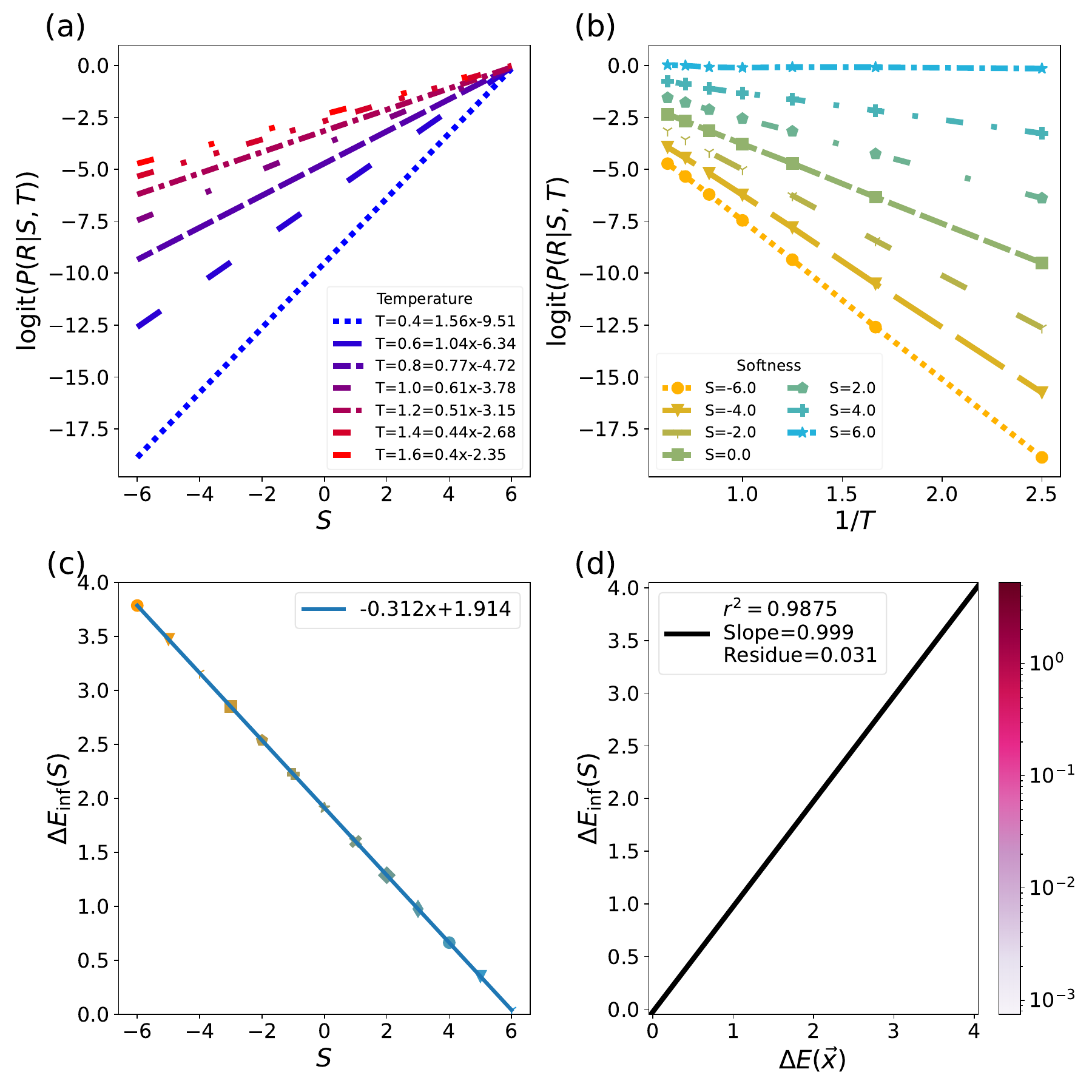}
    \caption{\textbf{Relationship between softness and $\Delta E (\vec x)$ for positive energy barriers and a balanced dataset.} The energy is sampled from a Gaussian distribution with the mean 2 standard deviations away from zero, and all negative energy samples are then discarded. Finally, we choose 50\% of samples where rearrangement was recorded, and 50\% where it was not. Plotting conventions are the same as in Fig.~\ref{fig:infinitedatalimit}. Note that the correct energy is learned (slope of 0.999), and the  spread in the 2D density plot is minimal.
    \label{fig:unsymmetricdata}}
\end{figure*}

Recall, as explained in the main text, that in the true system all energy barriers are positive. However, in the main text, we chose energy barriers to be symmetric around zero for simplicity. Figure~\ref{fig:unsymmetricdata} is the analogue of Fig.~\ref{fig:infinitedatalimit}, but now evaluated for all energy barriers being positive. We balance the training set, similarly to Ref.~\cite{Schoenholz2016}, so that the number of rearranging and non-rearranging particles is the same. The results are qualitatively unchanged from Fig.~\ref{fig:infinitedatalimit} in the main text. In particular, the correct energy barriers are learned. 

\label{appendunsymmetric}
 
\end{document}